\documentclass[12pt,preprint]{aastex}

\usepackage{CJK}
\usepackage{amsmath}
\usepackage{natbib}
\usepackage{longtable, lscape}
\usepackage{graphicx} 
\usepackage{epsfig}
\usepackage{xcolor}
\usepackage[colorlinks = true,linkcolor = blue,citecolor = blue]{hyperref}

\begin{document}
\begin{CJK}{UTF8}{gbsn}

\title{Evidence for Gas from a Disintegrating Extrasolar Asteroid\footnote{ The data presented herein were obtained at the W.M. Keck Observatory, which is operated as a scientific partnership among Caltech, the University of California and NASA. The Observatory was made possible by the generous financial support of the W.M. Keck Foundation.
}}

\author{S. Xu(许\CJKfamily{bsmi}偲\CJKfamily{gbsn}艺)\altaffilmark{a}, M. Jura\altaffilmark{b}, P. Dufour\altaffilmark{c}, B. Zuckerman\altaffilmark{b}}
\altaffiltext{a}{European Southern Observatory, Karl-Schwarzschild-Stra{\ss}e 2, 85748 Garching, Germany, sxu@eso.org}
\altaffiltext{b}{Department of Physics and Astronomy, University of California, Los Angeles CA 90095-1562; jura@astro.ucla.edu, ben@astro.ucla.edu}
\altaffiltext{c}{Institut de Recherche sur les Exoplan$\grave{e}$tes (iREx), Universit$\acute{e}$ de Montr$\acute{e}$al, Montr$\acute{e}$al, QC H3C 3J7, Canada; dufourpa@astro.umontreal.ca}

\begin{abstract}
We report high-resolution spectroscopic observations of \mbox{WD 1145+017} -- a white dwarf that recently has been found to be transitted by multiple asteroid-sized objects within its tidal radius. We have discovered numerous circumstellar absorption lines with linewidths of $\sim$ 300 km s$^{-1}$ from Mg, Ca, Ti, Cr, Mn, Fe and Ni, possibly from several gas streams produced by collisions among the actively disintegrating objects. The atmosphere of WD 1145+017 is polluted with 11 heavy elements, including O, Mg, Al, Si, Ca, Ti, V:, Cr, Mn, Fe and Ni. Evidently, we are witnessing the active disintegration and subsequent accretion of an extrasolar asteroid.

\end{abstract}


\section{Introduction}

Recent studies show that planetary systems are widespread around white dwarfs. About 25-50\% of white dwarfs show ``pollution" in their atmospheres, likely from accretion of planetesimals that were perturbed by planet(s) into the white dwarf's tidal radius \citep{Zuckerman2003, Zuckerman2010, Koester2014a}. High-resolution spectroscopic observations of these polluted white dwarfs typically reveal abundances similar to the observed compositions of solar system objects \citep{JuraYoung2014}. The most heavily polluted white dwarfs often have an infrared excess from an orbiting dust disk \citep{Kilic2006b}. Occasionally, circumstellar gaseous material is also detected, mostly via calcium infrared triplet emission \citep{Gaensicke2006, Melis2012, Debes2012b}.

These systems can be dynamically active. The infrared flux of the dust disk around WD J0959-0200 dropped $\sim$ 30\% within one year \citep{XuJura2014}. The gas disk around WD J1617+1620 was dissipated within a few years \citep{Wilson2014}. One model to explain these changes is the impact of an extrasolar asteroid onto a pre-existing dust disk \citep{Jura2008}. However, the parent body has ever been detected.

In this letter, we report Keck spectroscopic observations of a helium-dominated white dwarf WD 1145+017. This star was first discovered in the Hamburg/ESO survey \citep{Friedrich2000}. It caught our attention due to the strong infrared excess when comparing SDSS, UKIDSS and WISE photometry. After performing the Keck observations, we subsequently learned that this star happened to be located in a K2 field, and it was observed to be transitted by multiple objects with periods between 4.5 and 5.0 hr \citep{Vanderburg2015, Croll2015}. The system is rapidly evolving and the transit data are interpreted as cometary-like outflows from fragments of a larger parent body. Here, we provide additional evidence in support of this model.

\section{Observations and Data Reduction}

\subsection{HIRES}

WD 1145+017 was observed with the High Resolution Echelle Spectrometer (HIRES) \citep{Vogt1994} on the Keck I telescope on  2015 April 11 (UT). The blue collimator was chosen with the C5 deck, which has a slit width of 1{\farcs}148 and a resolution of $\sim$ 40,000. Three consecutive 2400 s exposures of WD 1145+017 were taken. The flux standard Feige 34 was observed with the same setup.

Data reduction was performed by using both MAKEE and IRAF following \citet{Klein2010, Xu2014}. We used the spectrum of Feige 34 to calibrate the target spectrum and reconstruct the profiles of broad helium lines. The three separate spectra were combined with equal weights to produce the final spectrum, which has a wavelength coverage from 3110 {\AA} to 5950 {\AA}. The signal-to-noise ratio (S/N) per pixel is typically over 25.

\begin{figure}[hp]
\plotone{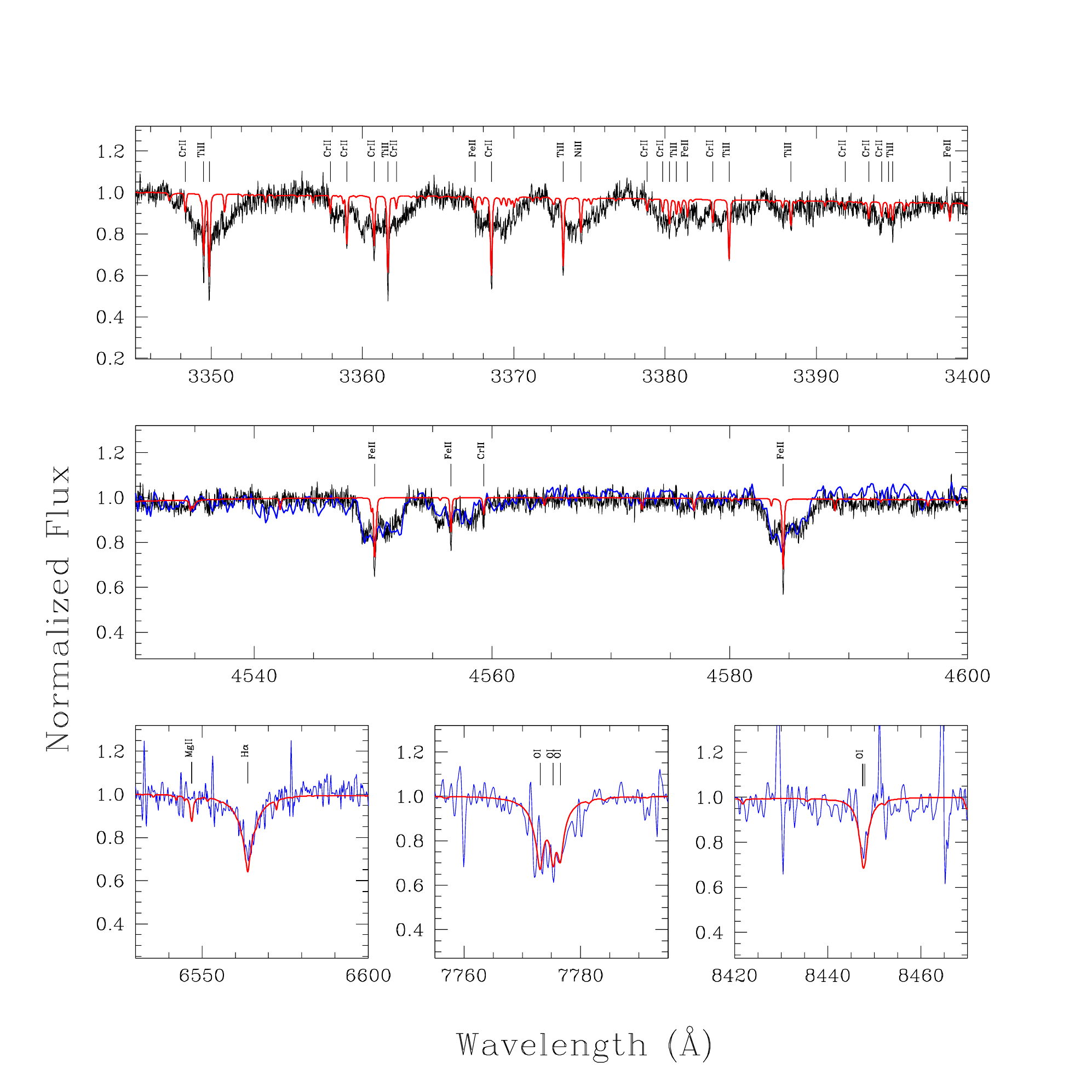}
\caption{Portion of the Keck spectra. The black and blue lines represent the HIRES and ESI data, respectively. The red line is our best fit model to the photospheric lines. They are all plotted in the heliocentric reference frame. We have detected numerous photospheric and circumstellar absorption lines.
\label{Fig1}}
\end{figure}

We identified $\sim$ 200 photospheric absorption lines from 15 different ions. About one third of these lines also show broad ($\sim$ 300 km s$^{-1}$) circumstellar absorption features, as shown in Figures \ref{Fig1} and \ref{Fig2}. All circumstellar lines arise from energy levels within 4.2 eV of the ground state (see Table \ref{Tab2}) . 

\subsection{ESI}

On 2015 April 25 (UT), we observed WD 1145+017 with the ESI (Echellette Spectrograph and Imager) \citep{Sheinis2002} on the Keck II telescope. The narrowest possible slit width of 0{\farcs}3 was chosen, corresponding to a resolution of $\sim$ 13,700. WD 1145+017 was observed with two consecutive 1180 s exposures. Data reduction was performed using the MAKEE package. Each echelle order was continuum normalized in IRAF with a low-order polynomial. The ESI spectrum has a continuous coverage from 3900 {\AA} to 10100 {\AA} with a S/N per pixel over 35. Some lines were only detected in the ESI data due to its coverage in the longer wavelength. Unfortunately, about two thirds of the photospheric lines lie shortward of 3900 {\AA}, which is not accessible with ESI. For the spectral region that overlaps with HIRES data (3900 - 5950 {\AA}), the agreement is quite good, as shown in Figures \ref{Fig1} and \ref{Fig2}. 

\section{Discussion}

\subsection{Photospheric Abundances \label{3.1}}

We adopted an effective temperature of 15,900 K and surface gravity of 8.0 for WD 1145+017 \citep{Vanderburg2015}. However, the surface gravity is highly uncertain because it cannot be derived unambiguously from spectroscopic fitting for helium white dwarfs below $\sim$ 16,500 K \citep{Voss2007}. To determine the abundances of heavy elements, we followed procedures outlined in \citet{Dufour2012} and fitted a spectral region of 10-15 {\AA} at a time. HIRES data were used primarily for the analysis due to its higher spectral resolution. The final abundances and the photospheric lines are shown in Table \ref{Tab1}. 

Oxygen only has lines in the lower resolution ESI data. The best fit model requires an oxygen abundance log n(O)/n(He) of -3.7 and -4.5 for the O I 7775 {\AA} triplet and O I 8446 {\AA}, respectively. The discrepancy cannot be due to contaminations from circumstellar absorption because the oxygen lines all arise from energy levels $\sim$ 9 eV above the ground state. A higher weight is given to O I 8446 {\AA} because the O I triplet is not resolved, as shown in Fig \ref{Fig1}.

The total mass of heavy elements currently in the outer convection zone of WD 1145+017 is 6.6 $\times$ 10$^{23}$ g, about 70\% the mass of Ceres. The planetesimal was dominated by four elements, O, Fe, Mg and Si, similar to rocky objects in the solar system and extrasolar asteroids accreted onto other polluted white dwarfs, as shown in Fig \ref{Fig3} \citep[e.g.][]{Klein2010, Dufour2012, JuraYoung2014}. The mass fraction of oxygen is $\sim$ 60\%, which is among the most oxygen-rich extrasolar planetesimals ever detected. However, due to the large error associated with the oxygen abundance, the oxygen abundance would be ``normal" if it is at the lower end. 

If the accretion is in a steady state \citep{Koester2009a}, the mass accretion rate for WD 1145+017 would be 4.3 $\times$ 10$^{10}$ g s$^{-1}$, among the highest of all polluted white dwarfs. The accretion needs to be on-going for 5 $\times$ 10$^5$ years to accumulate a total mass of 6.6 $\times$ 10$^{23}$ g. It is very unlikely that one tidal disruption event can last that long \citep{Veras2015}. In this case, the material currently in the white dwarf's photosphere would be associated with a previous tidal disruption event. Alternatively, WD 1145+017 could be in the very early stage of tidal disruption and it is experiencing a short high-accretion burst. Evidence for such bursts have been speculated in the literature but never been directly detected \citep{Farihi2012b, XuJura2014}. The mass accretion rate can go up to 10$^{13}$ g s$^{-1}$, approaching those of low accretion rate cataclysmic variables \citep{Schmidt2007}. In this scenario, the current tidal disruption event would have been on-going for 2000 yr. The heavy elements in WD 1145+017 could also come from a combination of the two scenarios.

\subsection{Circumstellar Gas}

A large number of circumstellar absorption lines are detected, as shown in Figures \ref{Fig1} and \ref{Fig2}. We are observing relatively cool circumstellar gas blocking part of the white dwarf's photosphere. The line profiles are broad, asymmetric and morphologically different from the narrow circumstellar features (that originate from ionization of the ISM) detected around hot white dwarfs \citep{Holberg1995, Dickinson2012}. They also differ from the symmetric blue-shifted Si IV absorption lines detected around PG 0843+517, which is attributed to the presence of hot circumstellar gas \citep{Gaensicke2012}. Here, we present some qualitative analysis on the gas around WD 1145+017.

The gravitational redshift is estimated to be 30 km s$^{-1}$ at the surface of WD 1145+017. In the heliocentric frame, the average radial velocity of the photospheric lines is 42 $\pm$ 2 km s$^{-1}$. We can derive the kinematic radial velocity of the white dwarf to be 12 km s$^{-1}$. In the reference frame of the white dwarf, the full velocity at zeroth intensity of the circumstellar absorption line extends from -100 to 210 km s$^{-1}$ (see Figure \ref{Fig2}). This broad line profile can be produced by several gas streams moving in non-circular motions and they are likely associated with the disintegrating planetesimals. Otherwise, if the gas was in a circular motion, circumstellar absorption lines would be narrow and only have a radial velocity of $\sim$ 12 km s$^{-1}$ -- the kinematic radial velocity of the white dwarf.  

As shown in Table \ref{Tab2}, most circumstellar lines have an equivalent width around 0.5 {\AA} even though the abundances and oscillator strengths vary by more than three orders of magnitude; the implication is that most of the circumstellar lines are optically thick. However, the lines of the trace constituent Mn might be optically thin because Mn II 3442.0 {\AA}\footnote{As shown in Figure \ref{Fig2}, there is also a nearby Fe I line at 3442.3 {\AA}. However, the Fe I line is unlikely to have circumstellar absorption due to its low oscillating strength (f=0.008).} and Mn II 3482.9 {\AA} arise from the same lower energy level but have different equivalent widths, which are in proportion to their oscillator strengths. We calculated a Mn II column density of 4.7 $\times$ 10$^{13}$ cm$^{-2}$. Assuming the circumstellar gas and the photosphere have the same elemental compositions, the Fe II column density N(Fe II) equals 3.7 $\times$ 10$^{16}$ cm$^{-2}$, leading to opaque lines. We can also estimate the excitation temperature, which is $\sim$ a few thousand Kelvins. The heating mechanism might be similar to the Z II region model described in \citet{Melis2010}.

As shown in Figure \ref{Fig2}, all the circumstellar lines have about the same depth, indicating $\sim$ 15\% of the white dwarf's photosphere is covered by gas, which has a vertical thickness h = 0.15 $\times$ 2R$_{wd}$ = 2700 km (R$_{wd}$ from \citet{Vanderburg2015}). If the radial extent of the gas goes to the tidal radius ($\sim$ 100 R$_{wd}$), the total mass of Fe II in the gas disk can be estimated as:

\begin{equation}
M(\textrm{Fe II})= 2 \pi \times 100R_{wd} \times h \times N(\textrm{Fe\  II}) \times m(\textrm{Fe}) 
\end{equation}

where m(Fe) is the mass of one Fe atom. The total mass of Fe II gas is 1.0 $\times$ 10$^{15}$ g. Assuming the mass fraction of Fe II is 25\%, the total gas mass M$_{gas}$ would be \mbox{4.2 $\times$ 10$^{15}$ g}. 

If gas is being accreted at V = 210 km s$^{-1}$ -- the maximum observed radial velocity-- we can derive a maximum accretion rate $\dot{M}$ as

\begin{equation}
\dot{M} =  M_{gas} \times \frac{V}{100R_{wd}} = 1.0 \times 10^{12} g\ s^{-1}
\end{equation}

This value is higher than the mass accretion rate onto the white dwarf's photosphere under the steady state assumption (4.3 $\times$ 10$^{10}$ g s$^{-1}$).  The accretion is likely to be a sporadic process. The gas lifetime can be calculated by dividing the total disk mass by the total accretion rate and it is only $\sim$ 1 hr. The gas must be constantly replenished. In addition, no variability was observed in the circumstellar absorption lines by comparing our 3 separate HIRES and 2 separate ESI observations, indicating a relatively smooth azimuthal distribution.

There are at least two ways to generate circumstellar gas. (i) Sublimation from the disrupting planetesimals \citep{Rappaport2012}. The planetesimal's surface can be heated to have a high-Z atmosphere and for planetesimals around WD 1145+017, the mass loss would be in a free streaming limit \citep{Vanderburg2015}. This scenario is very similar to the outgassing of a cometary nuclei when it gets close to the Sun \citep[e.g.][]{Keller1986}. However, the thermal speed is only $\sim$ 1 km s$^{-1}$ and it cannot explain the velocity dispersion of \mbox{300 km s$^{-1}$}. (ii) Collisions among the planetesimals. If the tidal disruption event resembles the encounter of comet Shoemaker-Levy 9 with Jupiter \citep{BennerMcKinnon1995}, there could exist a stream of disintegrating planetesimals in an eccentric orbit with the periastron inside the white dwarf's tidal radius. The observed velocity dispersion might represent different post-collision speeds. This is further supported by the broad and asymmetric circumstellar line profile. 

\subsection{Origin of the Infrared Excess}

WD 1145+017 belongs to a subgroup of polluted white dwarfs that show a strong infrared excess from a dust disk. However, the dust disk appears to be misaligned with the circumstellar gas and the transiting planetesimals \citep{Vanderburg2015}. Here, we discuss several possibilities.

{\it A puffed-up dust disk}. We assume the dust particles at the outer edge 100R$_{wd}$ have a vertical speed of 30 km s$^{-1}$, a small fraction of the observed spread in the radial velocity of the circumstellar gas. The dust can travel to a vertical heigh of 10 R$_{wd}$, which is sufficient to produce the infrared excess.

{\it A precessing dust disk}. At least one more object (very likely a planet) is required to be present in the system to perturb the planetesimal into the white dwarf's tidal radius. If the orbital plane of the planet and the dust disk are not aligned, the planet could induce significant precession on the dust disk. In this case, we would expect the infrared flux to vary gradually.

{\it A previous generation dust disk}. The dust disk can be produced by a previous tidal disruption event, possibly associated with the one that also polluted the white dwarf's atmosphere. The lifetime of a dust disk is estimated to be between 10$^5$ to 10$^7$ yr \citep[e.g.][]{Barber2012}. \citet{Jura2008} suggested that a new coming object would have a different orbital inclination and mutual collisions can lead to partial evaporation or even total disruption of the dust disk. This scenario is further supported by the discovery of rapid changes in the dust and gas disk around polluted white dwarfs \citep{XuJura2014, Wilson2014}. WD 1145+017 could be a precursor to these systems.

\section{Conclusions}

WD 1145+017 is in a unique stage that has a polluted atmosphere, a dust disk, cool circumstellar gas and multiple transiting planetesimals. In conclusion:

\begin{itemize}

\item We have detected 11 heavy elements in the photosphere. The composition of the accreted planetesimal is similar to those observed in other polluted white dwarfs. The heavy elements could either come from a burst of accretion induced by the disintegrating planetesimals, a previous tidal disruption event, or both.

\item We have uniquely detected $\sim$ 70 circumstellar absorption lines from 8 ions. They have very broad asymmetric profiles with velocity dispersion $\sim$ 300 km s$^{-1}$. The gas lifetime is very short and must be constantly replenished, possibly by collisions among the disrupting planetesimals. 

\item The infrared excess could come from either a puffed-up dust disk, a precessing dust disk or a previous generation of dust disk.

\end{itemize}

We thank N. Mahesh for helping with the HIRES observing run.  This work has been partly supported by the NSF. The authors wish to recognize and acknowledge the very significant cultural role and reverence that the summit of Mauna Kea has always had within the indigenous Hawaiian community.  We are most fortunate to have the opportunity to conduct observations from this mountain.

\end{CJK}

\newpage

\begin{table}[hp]
\begin{center}
\caption{Photospheric absorption lines and abundance determination}
\begin{tabular}{lllllll}
\\
\hline \hline
Ion & log n(Z)/n(He)	& t$^a$	& M$^{b}$	&$\dot{M}$$^{c}$ \\
	&	&	(10$^5$ yr)	& (10$^{20}$ g)	& ( 10$^8$ g s$^{-1}$)\\
  \hline

H			& -4.7 $\pm$ 0.1 & ...	& ...	& ...\\
C		 	& $<$ -4.3 	& 6.0	& $<$ 3200	& 	$<$ 170 \\
O			& -4.3 $\pm$ 0.5 & 5.5		& 4300	& 	240 \\
Mg		& -5.49 $\pm$ 0.13 & 5.5	& 410	&	24\\
Al 		 & -6.74 $\pm$ 0.14 & 5.2	&26	& 1.6\\
Si 		& -5.69 $\pm$ 0.09 &	5.0	& 300	& 19\\
Ca	&  -6.57 $\pm$ 0.12 & 3.6	&58	&	5.1\\
Ti 		& -8.04 $\pm$ 0.19 &  3.4	& 2.3	&	0.22\\
V:		& -8.7:	&3.4	& 0.57	&	0.054\\
Cr 	  & -7.31 $\pm$ 0.12 &  3.5	& 14	& 1.2 \\
Mn		& -8.25 $\pm$ 0.20 &3.5	& 1.7	&	0.15\\
Fe			&	-5.35 $\pm$ 0.11 &3.6	&1300	&	120 \\
Ni			& -6.24 $\pm$ 0.18 & 3.5	& 180 &	16\\
total	&	& & 6600	& 430 \\
\hline
\label{Tab1}
\end{tabular}
\end{center}
\end{table}

{\bf Notes.} \\
$^a$ Diffusion time out of the convection zone, following \citet{Dufour2012}. \\
$^{b}$ Current mass in the white dwarf's convection zone. \\
$^{c}$ Accretion rate assuming a steady state \citep{Koester2009a}. \\
\indent The following lines are free from circumstellar absorption and used for abundance determinations: H I 4861.3 {\AA}, 6562.8 {\AA}; C I 4267.3 {\AA}; O I  7775 {\AA}$^d$, 8446.4 {\AA}$^d$; Mg I 3829.4 {\AA}, 5172.7 {\AA}, 5183.6 {\AA}; Mg II 4481.1 {\AA}$^e$, 7877.1 {\AA}$^d$, 7896.0 {\AA}$^d$, 7896.4 {\AA}$^d$; Al II 3586.5 {\AA}, 3587.0 {\AA}, 3961.5 {\AA}, 4663.1 {\AA}; Si II 3853.7 {\AA}, 3856.0 {\AA}, 3862.6 {\AA}, 4128.1  {\AA}, 4130.9  {\AA}, 5041.0  {\AA}, 5056.0  {\AA}, 6347.1  {\AA}, \mbox{6371.4  {\AA}}; Ca II 3158.97 {\AA}, 3181.37 {\AA}, 3706.07 {\AA}, 3736.97 {\AA}, 8498.07 {\AA}$^d$, 8542.17 {\AA}$^d$; Ti II 3239.0 {\AA}, 3261.6 {\AA},3329.5 {\AA}, 3504.9 {\AA}; V II, 3271.1 {\AA}; Cr II 3125.0 {\AA}, 3180.7 {\AA}, 3217.4 {\AA}; Mn II, 3482.9 {\AA}; Fe I, 3165.9 {\AA}, 3565.4 {\AA}, 3570.1 {\AA}, 3606.7 {\AA}, 3608.9 {\AA}, 3618.8 {\AA}, 3631.5 {\AA}, 3719.9 {\AA}, \mbox{3721.6 {\AA}}, 3749.5 {\AA}, 3758.2 {\AA}, 3815.8 {\AA},3820.4 {\AA}, 3825.9 {\AA}, 3859.9 {\AA}, 4063.6 {\AA}, 4071.7 {\AA}, 4307.9 {\AA}; Fe II, 3144.8 {\AA}, 3177.5 {\AA}, 3180.2 {\AA}, 3183.1 {\AA}, 3186.7 {\AA}, 3237.8 {\AA}, 3266.9 {\AA}, 4635.3 {\AA}, \mbox{4923.9 {\AA}}, 5260.3 {\AA}, 6456.4 {\AA}$^d$; Ni I, 3524.5 {\AA}; Ni II, 3374.0 {\AA}, 3471.4 {\AA}. ($^d$ This line is only detected in the ESI data and is given less weight. $^e$ This line is blended. ) \\

\clearpage

\begin{figure}[hp]
\plotone{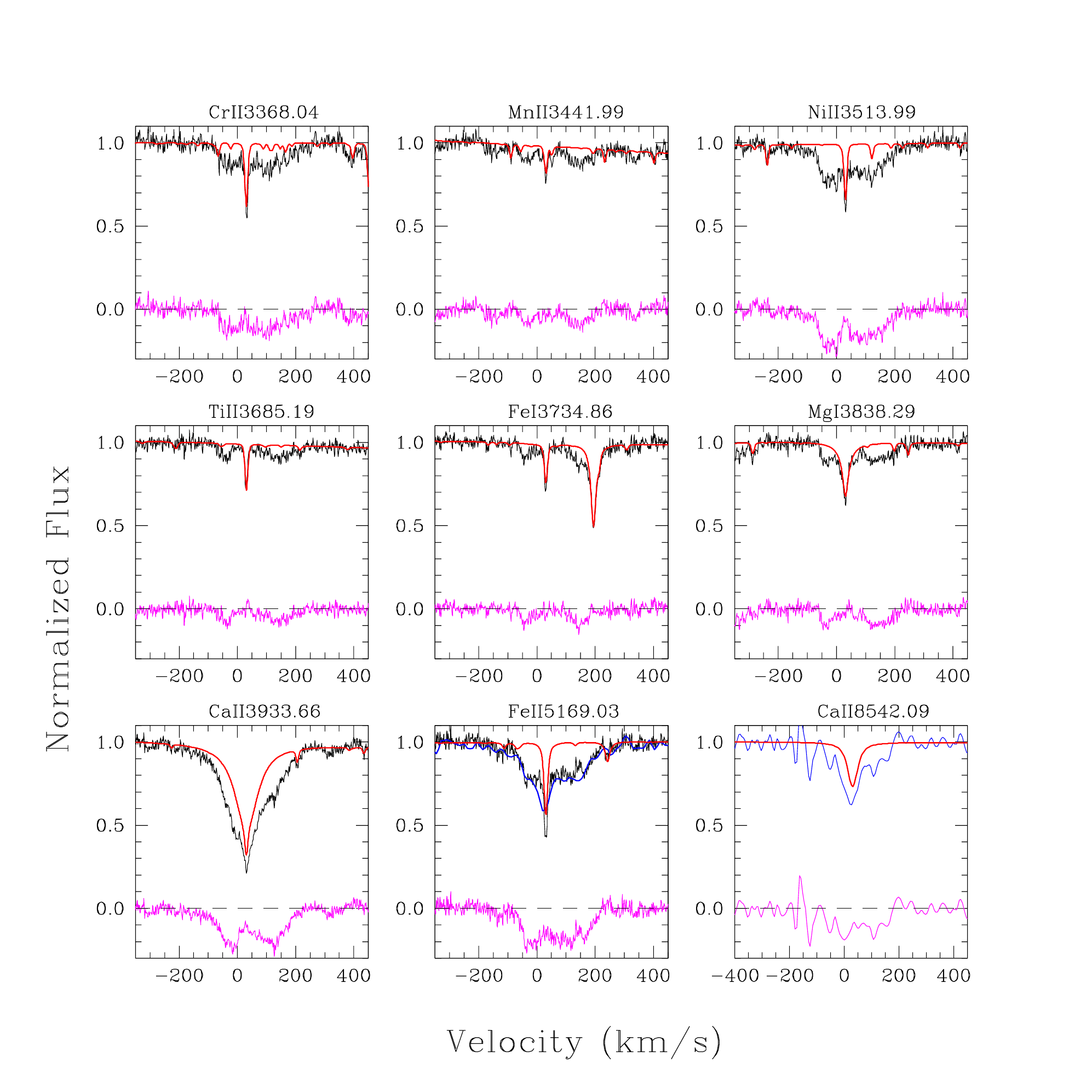}
\caption{Similar to Figure \ref{Fig1} except in velocity space in the white dwarf's reference frame. The magenta line represents data minus model -- there exists circumstellar absorption produced by the gas between us and the white dwarf. All the photospheric lines have a velocity of 30 km s$^{-1}$ due to the gravitational redshift. The circumstellar lines are very asymmetric and extend from -100 km s$^{-1}$ to 210 km s$^{-1}$.
}
\label{Fig2}
\end{figure}

\begin{figure}[hp]
\plotone{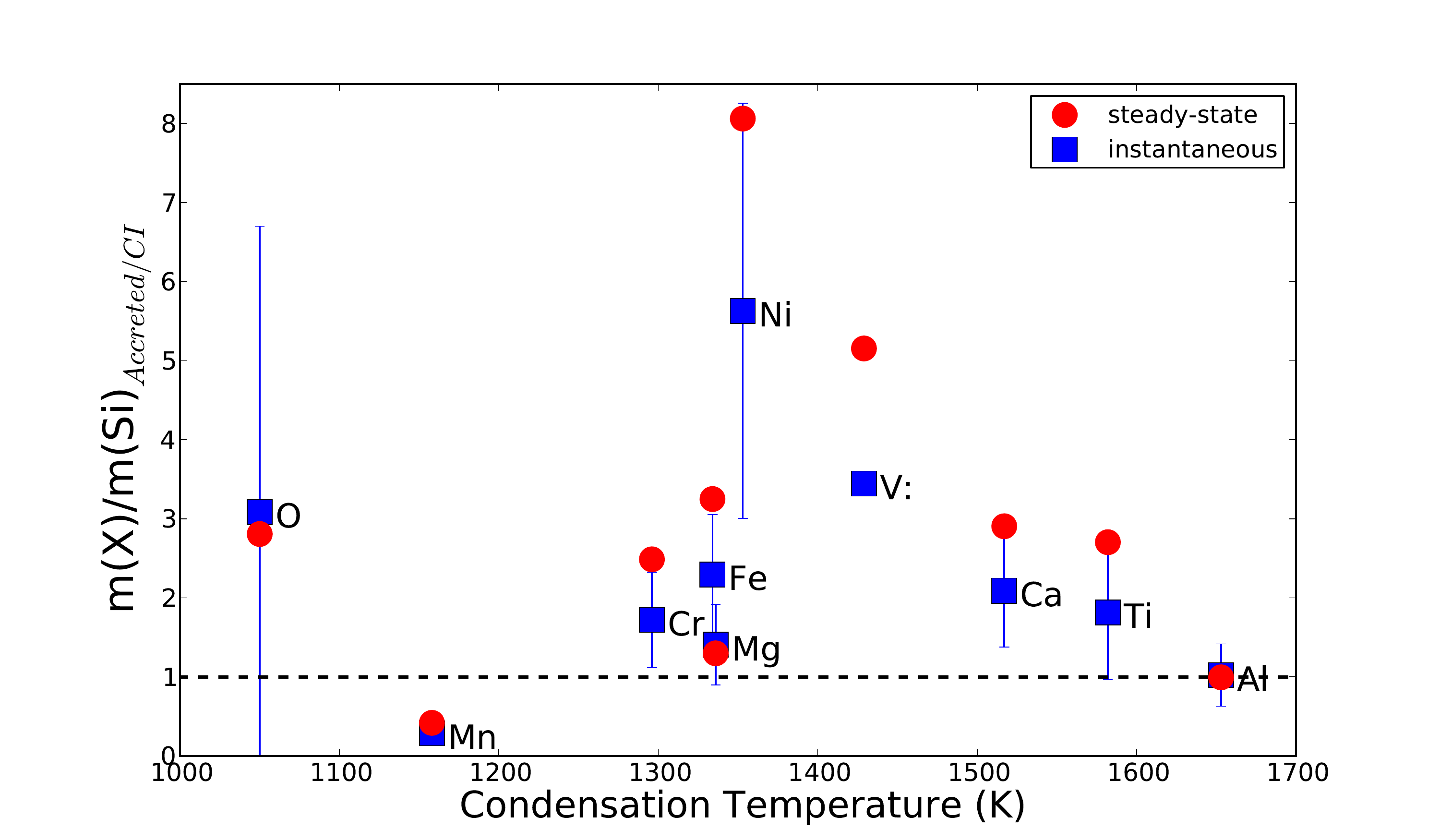}
\caption{The mass fraction of an element relative to Si in the extrasolar asteroid accreted onto WD 1145+017. The values are normalized to the ratios in CI chondrites \citep{WassonKallemeyn1988}. For example, a CI-chondrite analog would lie along the black dashed line. Two models are presented and the error bar is only shown for the instantaneous model. Within the uncertainties, the overall abundance pattern in the object accreted onto WD 1145+017 is similar to that in CI chondrites. 
}
\label{Fig3}
\end{figure}

\begin{center}
\begin{longtable}{lllllll}
\caption{Circumstellar lines in WD 1145+017}
\\
\hline \hline
Ion & $\lambda^a$	& E$_{low}$ & f	& EW$^b$	& N$^c$\\
	& ({\AA})	& (eV)	&	& ({\AA})	& (10$^{12}$ cm$^{-2}$) \\
  \hline
\endfirsthead 

\multicolumn{6}{c}{Table \ref{Tab2} --- \emph{Continued}} \\ 
\hline
\hline
Ion & $\lambda^a$	& E$_{low}$ & f	& EW$^b$	& N$^c$\\
	& ({\AA})	& (eV)	&	& ({\AA})	& (10$^{12}$ cm$^{-2}$) \\
\hline
\endhead

\endfoot
  
 \hline
\endlastfoot
Mg I	& 3832.3	& 	2.7	&	0.45	& 0.05	& $>$ 0.9 \\		
Mg I	&	3838.3	& 2.7	&	0.5	& 0.2	& $>$ 3 \\		

Ca II$^d$	&	3933.7	& 0.0	&	0.68	&	1.0	& $>$11\\		
Ca II$^d$	&	3968.5	& 0.0	&	0.33	&	0.8	 & $>$ 17\\	
Ca II$^d$		&	8542.1	& 1.7	&	0.02		& 0.2	& $>$ 18	\\

Ti II		& 3341.9	&	0.6		&0.38	&0.3	&	$>$8 \\		
Ti II		& 3685.2	&	0.6		&0.13	&0.2	&	$>$10 \\		
Ti II		& 3761.3	&	0.6		&0.26	&0.1	&	$>$3	\\	
Ti II		& 3759.3	&	0.6		&0.2	&0.13	&	$>$5	\\	

Cr II	& 3125.0	& 	2.5		& 0.31		&0.5	& $>$17	\\	
Cr II	& 3132.1	& 	2.5		& 0.33		&0.4	& $>$15	\\	
Cr II	& 3368.1	&	2.5		& 0.10		&0.3	& $>$31	\\	
Cr II	& 3403.3	&	2.4		& 0.043		&0.2	& $>$46	\\	
Cr II	& 3408.8	&	2.5		& 0.051		&0.5	& $>$96	\\	
Cr II	& 3422.7	&	2.5		& 0.064		&0.3	& $>$51	\\	
Cr II	& 3433.3	&	2.4		& 0.047		&0.2	& $>$43	\\	

Mn II	& 3442.0	& 1.8	&0.05	& 0.2	 & $>$46 \\		
Mn II	& 3482.9	& 1.8	&0.029	& $<$ 0.15	& $<$48 \\	

Fe I		& 3581.2	& 0.9	&0.23	&	0.1		& $>$ 4.6 \\		
Fe I		& 3734.9	& 0.9	&0.19	&	0.2		& $>$ 8.6	\\

Fe II	& 3281.3	&	1.0	& 0.0004	& 0.4	& $>$9900 \\	
Fe II	& 3295.8	&	1.1	& 0.0003	& 0.4	& $>$13000 \\		
Fe II	& 3227.7	&	1.7	& 0.018		& 0.7	& $>$390 \\
Fe II	& 4173.5	&	2.6	& 0.0012	& 0.2	& $>$840 \\	
Fe II	& 4178.9	&	2.6	& 0.0006	& 0.2	& $>$2000 \\	
Fe II	& 4233.2	&	2.6	& 0.0018	& 0.5	& $>$1600 \\		
Fe II	& 4303.2	&	2.7	& 0.0006	& 0.2	& $>$1900 \\		
Fe II	& 4351.8	&	2.7	& 0.0014	& 0.4	& $>$1500 \\		
Fe II	& 4583.8	&	2.8	& 0.0012	& 0.6	& $>$2700 \\	
Fe II	& 4629.3	&	2.8	& 0.0005	& 0.2	& $>$2700 \\	
Fe II	& 4555.9	&	2.8	& 0.0005	& 0.3	& $>$3100 \\		
Fe II	& 4515.3	&	2.8	& 0.0007	& 0.2	& $>$1500 \\	
Fe II	& 4522.6	&	2.8	& 0.0017	& 0.4	& $>$1300 \\		
Fe II	& 4508.3	&	2.9	& 0.0009	& 0.3	& $>$1700 \\		
Fe II	& 5169.0	&	2.9	& 0.017		& 0.8	& $>$200 \\	
Fe II	& 4549.5	&	2.9	& 0.001		& 0.5	& $>$2700 \\		
Fe II	& 5316.6	&	3.2	& 0.0002	& 0.5	& $>$9100 \\		
Fe II	& 5276.0	&	3.2	& 0.0013	& 0.4	& $>$1100 \\		
Fe II	& 5234.6	&	3.2	& 0.0008	& 0.3	& $>$1600 \\		
Fe II	& 5197.6	&	3.2	& 0.0011	& 0.3	& $>$1100 \\		
Fe II	& 3154.2	&	3.8	& 0.031		& 0.5	& $>$180 \\	
Fe II	& 3167.9	&	3.8	& 0.0024	& 0.6	& $>$280 \\		
Fe II	& 3289.4	&	3.8	& 0.0034	& 0.3	& $>$770 \\		
Fe II	& 3135.4	&	3.9	& 0.013		& 0.5	& $>$420 \\		
Fe II	& 3323.1	&	4.0	& 0.003		& 0.4	& $>$1400 \\		
Fe II	& 3162.8	&	4.2	& 0.008		& 0.7	& $>$920 \\	
Fe II	& 3493.5	&	4.2	& 0.0059	& 0.4	& $>$670 \\		

Ni II		& 3513.9	& 0.2	& 0.0033	& 0.5	& $>$1400 \\
Ni II	& 3769.5	& 3.1	& 0.0036	& 0.1	& $>$280 \\	
Ni II	& 3576.8	& 3.1	& 0.0059	& 0.3	& $>$510 \\		
Ni II	& 4067.0	& 4.0	& 0.0015	& 0.1	& $>$370 \\		
\label{Tab2}
\end{longtable}
\end{center}
{\bf Notes.} This is a selection of relatively strong and unblended circumstellar lines. For each ion, they are arranged in terms of increasing lower energy level.\\
$^a$ Wavelength is listed in air. \\
$^b$ The equivalent width is measured from a direct integral of the flux below the continuum. This value is usually dominated by the circumstellar line because the strength of the photospheric line is much weaker.  \\
$^c$ N is the column density of the ion, following equation 3-48 in \citet{Spitzer1978}. \\
$^d$ The observed Ca lines are much stronger than the best-fit model considering all Ca lines. We conclude that they have a circumstellar contribution. The equivalent width is measured from the spectra that excludes photospheric Ca line, i.e. the magenta line Figure \ref{Fig2}.

\clearpage
\newpage

\clearpage
\newpage
\bibliographystyle{apj}
%\bibliography{apj-jour,WD.bib}

\end{document}